# Interface states in the rectangular lattice photonic crystals with identical dielectric rods

**Jianzhi Chen, Jianlan Xie, Exian Liu, Bei Yan, Jianjun Liu***

Key Laboratory for Micro/Nano Optoelectronic Devices of Ministry of Education & Hunan Provincial Key Laboratory of Low-Dimensional Structural Physics and Devices, School of Physics and Electronics, Hunan University, Changsha 410082, China

*E-mail address: jianjun.liu@hnu.edu.cn

**Abstract:** Interface states in photonic crystals (PCs) have attracted attention for the special properties, such as high transmission efficiency in bend waveguides, and their generation related to the topological phase. Previous works on interface states in PCs were mainly based on the square lattice, the honeycomb lattice or the triangular lattice, but with different materials, shapes, or sizes of rods resulting in the complicated structure. In this paper, an interface is constructed by two 2D PCs with different rectangular lattices, but the same materials, shapes, and sizes of dielectric rods, which generates interface states. The interface states are analyzed with respect to Zak phases and surface impedances. The retainability of the interface states in rectangular lattice PCs is investigated by studying the relationship between the length-width ratio of the rectangular lattice and the Zak phase. It is found that, when the interface states are generated by changing the length-width ratio of the rectangular lattice, the retainability of the interface states is related to the positions of the photonic bandgaps or the Zak phases of the bands. A more detailed examination indicates that these conclusions are applicable to the rectangular lattice PCs with other materials, shapes, and sizes of dielectric rods. These results can lead to new ways to generate interface states easily, with only one kind of dielectric rod. In addition, these outcomes may contribute to the understanding of the relationship between the geometry and the interface state.

1. **Introduction**

PCs are artificial periodic [1,2] or quasi-periodic [3] structured materials with different permittivities. Due to the properties of their photonic bandgaps (PBGs), photonic localization, negative refraction, and others, PCs can be used for many purposes, such as optical fibers [4–14], lenses [15–26], prisms [27,28], sensors [11, 13,29–32], lasers [33–35], filters [8,36], one-way waveguides [37–40], and even logic gates [41], where waveguides are the primary elements of integrated optics [42] that can control the direction of light propagation.

Photonic crystal waveguides (PCWs) can control the light propagation at a specific frequency according to the properties associated with the PBG and the waveguide mode [43,44]. The transmission loss of such a PCW is low [45]. The interface states in PCs can be used for both photonic localization and surface mode coupling [46–54], which allows the control of light propagation and the utilization of waveguiding effect. Nearly 100% transmission can be achieved by a bend waveguide with interface states [55]. Moreover, other special properties of interface states have attracted the attention of researchers. For example, the existence of interface states depends on the PBGs, the surface impedances, and the Zak phases of the two PCs [56–65]. Due to topological protection, interface states can exist between the two PCs with different topological properties [66–68], and the Zak phase is an important topological property [57–64]. Topological properties [69,70] increase the usability of PCs, which will be explained in more detail below. The methods that are commonly used to generate interface states in PCs include band inversion [59–61], structure inversion [62,63], and unit-cell translation [64]. In addition, the generation of interface states can be analyzed only with Zak phases and surface impedances [65]. In PCs, due to the strong effect of the lattice on the bands, any change of the lattice structure may affect the generation and the properties of the interface states. Previous studies on interface states of PCs were mainly based on the square lattice [59–65], the honeycomb lattice [59] or the triangular lattice [62]. Because the square lattice is only a special case of the rectangular lattices, it is meaningful to study the characteristics of the interface states in rectangular lattice PCs in more depth, which contributes to obtain the more universal characteristics of interface states. In addition, according to previous research on interface states in PCs, the materials [59–63,65], the shapes [62,64], and the sizes [59–62,65] of the scatterers of the two PCs making up the interface, are different, which increases the difficulty of both preparation and commercialization. Therefore, it would be important to find a method to generate interface states with identical dielectric rods on both sides of the interface.

In order to generate interface states with identical dielectric rods on both sides of the interface and learn about the properties of the interface states in rectangular lattice PCs, the PBG, which is related to the generation of the interface states in PCs, will be analyzed with classification and comparison in this paper. Interface states can be generated with the same PBGs (e.g., the $m^{\text{th}}$ PBG of $PC_1$ and the $n^{\text{th}}$ PBG of $PC_2$, $m=n$) or the different PBGs (e.g., the $m^{\text{th}}$ PBG of $PC_1$ and the $n^{\text{th}}$ PBG of $PC_2$, $m \neq n$) of the two PCs, while the PBGs are related to the geometry. By summarizing previous works including band inversion [59–61], structure inversion [62,63], and unit-cell translation [64], it can be found that, firstly, if the interface states are generated via band inversion, the interface states are usually generated with the different PBGs. However, this method requires the formation of a Dirac point first. In other words, a careful selection of materials and structures is required. Secondly, if the interface states are generated via structure inversion, due to the limitation of the relationship between the geometric structure and the band structure, the interface states exist, generally, with the same PBGs. Finally, for unit-cell translation, the generation of the interface states requires the same PBGs, due to the invariability of the bands. As mentioned above, the methods that are commonly used to generate interface states in PCs are related to the positions of the PBGs of the two PCs. It can be assumed that the characters of the interface states are related to the positions of the PBGs of the two PCs, which can be affected by the lattices. When the length-width ratio of a rectangular lattice is used as the only variant to study the interface states, despite the difference from previous methods, it is still possible to generate interface states by changing the positions of the PBGs and the topological properties of the bands. In detail, the difference between the previous methods and the method in this paper is that, though deterministic interface states can be generated with the previous methods, it is convenient to control the realization, disappearance and optimization of the waveguide with the interface states with identical dielectric rods on both sides of the interface in rectangular lattice PCs. The study of the interface states with rectangular lattices can provide a new way to generate interface states and lead to understand their characteristics by comparison (Interface states generated with the same PBGs and the different PBGs).

In this paper, an interface is constructed with the two PCs using different rectangular lattices but the same materials, shapes, and sizes of the dielectric rods. This approach generates interface states more easily because of the identical dielectric rods. The interface states are generated by changing the positions of the PBGs and the topological properties of the bands, and it is found that the retainability (which is defined below to research the properties of interface states) of the interface states with rectangular lattices is related to the positions of the PBGs or the Zak phases of the bands. The similar results can be obtained when changing the materials, the shapes, and the sizes of the dielectric rods.

## 2. Model and theory

The PCW model, which we use in this paper to generate interface states, is shown in Fig. 1. The waveguide structure consists of two semi-infinite rectangular-lattice PCs. The radii of the circular dielectric rods of the two PCs are $r = 0.2$ μm, the materials of the rods are silicon with the relative permittivity $\varepsilon_r = 11.9$, and the background material is air. The lattice constants of the two PCs are set as $a_x = a = 1$ μm in the $x$ direction, and $a_{y1}$ (in $PC_1$) and $a_{y2}$ (in $PC_2$) (both are referred to as $a_y$) in the $y$ direction respectively. The ratio of the lattice constants in the two directions is defined as the length-width ratio of the rectangular lattice $\eta = a_y/a_x$. The generation and properties of interface states are studied by changing $a_{y1}$ and $a_{y2}$, i.e., by changing the length-width ratios of the rectangular lattices of the two PCs.

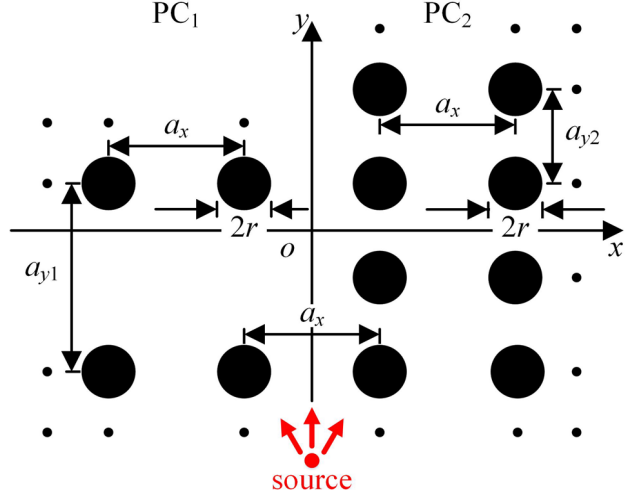

Fig. 1. The PCW model to generate interface states along the $y$ axis. The point light-source (red point) is located at the $y$ axis below the model.

For the 2D PCs, the necessary condition to generate an interface state is that, at a specific frequency with a specific $k_y$ (in a 1D system), the sum of the surface impedances ($Z$) on both sides of the interface is 0 [71–74]. Within the range of the PBG, the real part of the surface impedance (Re($Z$)) of PC is 0, and only the imaginary part of the surface impedance (Im($Z$)) affects how the wave reflects and propagates at the interface. In other words, to generate an interface state, it is sufficient that only the sum of the imaginary parts of the surface impedances is 0. Therefore, the necessary condition is

$$\text{Im}(Z_1) + \text{Im}(Z_2) = 0, \tag{1}$$

where $Z_1$ ($Z_2$) is the surface impedance of PC$_1$ (PC$_2$) for a specific $k_y$ and a specific frequency. In a 1D system (e.g., a 1D PC, or a 2D PC at a specific $k$) with mirror symmetry, the relationship between Im($Z$) of the adjacent PBGs (here, we define the $(n-1)^{th}$ PBG as $gap(n-1)$ and the $n^{th}$ PBG as $gap(n)$) and the Zak phase ($\varphi$)

of the $(n\text{-}1)^{\text{th}}$ band [60–64] is

$$\frac{\text{sgn}[\text{Im}(Z_{gap(n)})]}{\text{sgn}[\text{Im}(Z_{gap(n-1)})]} = -\exp(i\varphi_{n-1}). \quad (2)$$

Using Eq. (2), the signs of Im($Z$) can be obtained according to the Zak phases ($\varphi$), and subsequently the existence of interface states can be analyzed with Eq. (1).

## 3. Results and discussion

To simplify the study, the band of interface states is calculated under the condition with $a_{y1} = 2a_{y2}$ first, and then the characteristics of the interface states are studied with invariable $a_{y1}$ and changing $a_{y2}$ under two different premises, as shown in 3.1 and 3.2. The comparison between the interface states in 3.1 and 3.2 is discussed to probe into the factors of the generation of the interface states in rectangular lattice PCs, as shown in 3.3.

*3.1. Generation and description of the interface states with the different PBGs of the two PCs*

If $a_{y1} = 1.4$ μm ($\eta_1 = 1.4$), $a_{y2} = 0.7$ μm ($\eta_2 = 0.7$), the projected band structures for TM polarization (the below results are all for TM polarization) of the two PCs and the corresponding eigenfield distribution are respectively shown in Fig. 2(a) and 2(b). In Fig. 2(a), the interface states exist in the common PBG. According to Fig. 2(b), the eigenfield distribution, which is attained according to the calculation of bands in Fig. 2(a), is localized at the interface of the two PCs.

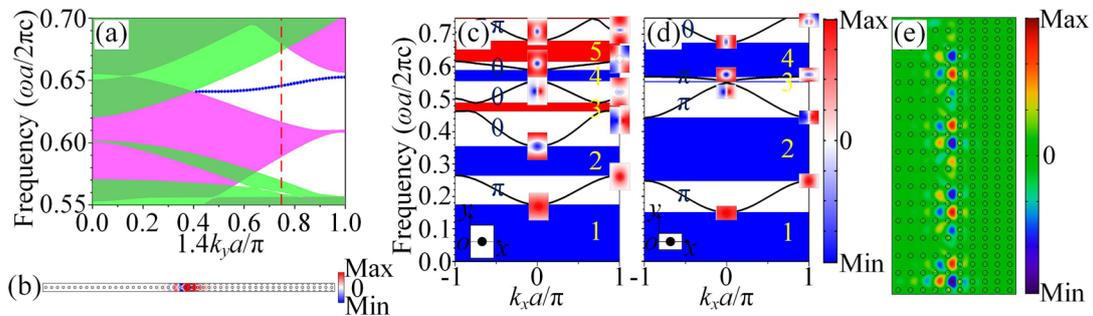

Fig. 2. (a) For $a_{y1} = 1.4a$ and $a_{y2} = 0.7a$, the projected band structures of the two PCs (the red dashed line indicates

$k_y = 3\pi/(5.6a)$). The pink regions represent the passing bands of PC$_1$, and the light-green regions represent the passing bands of PC$_2$. The dark-green regions represent the common bands of the two PCs. The band marked with a blue dotted line represents the band of the interface states; (b) the eigenfield distribution of the interface state at $f = 0.6457c/a$ with $k_y = 3\pi/(5.6a)$ in the projected band structures (the numbers of dielectric rods in the $x$ direction for the regions of both PC$_1$ and PC$_2$ are 25); the projected band structures in the $k_x$ direction, eigenfield distributions, and Zak phases at $k_y = 3\pi/(5.6a)$ of the two PCs: (c) PC$_1$; (d) PC$_2$. The colors of the PBGs represent the signs of Im($Z$), with red indicating positive and blue indicating negative. The coordinates for calculating surface impedances and Zak phases are shown in the lower left corner of the diagrams, and the origins are in the center of the left boundaries of the cells. The ordinal numbers of the PBGs are labeled with yellow; (e) the field distribution of the generated interface state at $f = 0.6457c/a$ (the numbers of dielectric rods in the $x$ direction for the regions of both PC$_1$ and PC$_2$ are 5, and the numbers of dielectric rods in the $y$ direction for the regions of PC$_1$ and PC$_2$ are 16 and 32, respectively).

Referring to the method of the calculation of Zak phases in a 1D system, the projected band structures in the $k_x$ direction, the eigenfield distributions, and the Zak phases at $k_y = 3\pi/(5.6a)$ of the two PCs (PC$_1$ and PC$_2$) are respectively shown in Fig. 2(c) and 2(d).

The coordinate system is established in the cell, and the origin is located at the edge of the cell because the surface impedance and the Zak phase are calculated at the edge of the cell [62], as shown in the lower left corner of Fig. 2(c) and 2(d). Considering the symmetry of the eigenmodes at the two high symmetric points in the reduced 1D Brillouin zone, the Zak phase of each band can be obtained, with the two highly symmetric points $k_x = -\pi/a$ (labeled by A) and $k_x = 0$ (labeled by B). According to Kohn's results [75], the Zak phase is $\pi$ if one of $|\tilde{E}_A(x=0)|$ and $|\tilde{E}_B(x=0)|$ is 0 and the other is not 0. In addition, the Zak phase is 0 if both are 0 or neither is. Therefore, according to the eigenfield distributions, the Zak phase corresponding to each band can be obtained—see Fig. 2(c) and 2(d). Here, $\tilde{E}_{\vec{k}}(x=0) = (1/a_y)\int_0^{a_y} E_{\vec{k}}(0,y)e^{-ik_y \cdot y}dy$, where $\vec{k}$ is relative to point A or B. Because Im($Z$) of the lowest PBG is always negative [60], according to Eq. (2), the sign of Im($Z$) for each PBG can be obtained—see Fig. 2(c) and 2(d). When $f \in [0.6165c/a, 0.6734c/a]$, Im($Z_1$) > 0, and Im($Z_2$) < 0 (where the fifth PBG of PC$_1$ and the fourth PBG of PC$_2$ overlap). In the range of PBG, as the frequency increases, Im($Z$) decreases from +∞ to 0 when Im($Z$) > 0. On the other hand, Im($Z$) decreases from 0 to -∞ when Im($Z$) < 0 [60]. Therefore, there is a frequency that can make Im($Z_1$) and Im($Z_2$) satisfy Eq. (1) within the frequency range where the two PBGs overlap. In other words, an interface state exists at the interface of the two PCs at this frequency [61]. These calculation results are obtained by finite element method.

The field distribution of the generated interface state at $f = 0.6457c/a$, selected from Fig. 2(a) and corresponding to Fig. 2(b), is shown in Fig. 2(e). An interface state can be observed, and the electric field is localized at the interface between the two PCs.

In order to explore the effect of rectangular lattice PCs on interface states, the length-width ratio ($\eta_2$) of the rectangular lattice of PC$_2$ is changed below (PC$_1$ is invariable), while the retainability (whose definition is below) of interface states is studied under the premise of generating interface states with the different PBGs (the fifth PBG of PC$_1$ and the fourth PBG of PC$_2$) at $k_y = 3\pi/(5.6a)$. In addition, Zak phases and surface impedances are analyzed. The results are shown in Fig. 3.

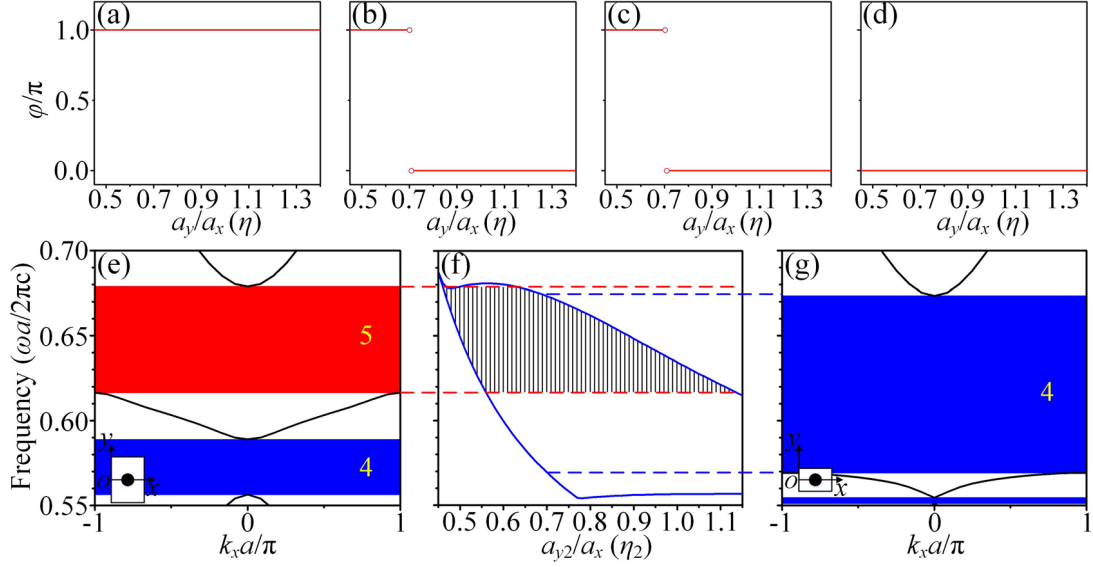

Fig. 3. (a)–(d) For $k_y = 3\pi/(5.6a)$ and $a_y \in [0.45a, 1.4a]$, the Zak phase is shown for: (a) the first band, (b) the second band, (c) the third band, (d) the fourth band; (e) when $a_{y1} = 1.4a$ and $k_y = 3\pi/(5.6a)$, the projected band structures intercepted from Fig. 2(c) along the $k_x$ direction. The ordinal numbers of the PBGs are labeled with yellow; (f) when $a_{y1} = 1.4a$, $a_{y2} \in [0.45a, 1.15a]$ and $k_y = 3\pi/(5.6a)$, the schematic diagram of the frequency ranges of the corresponding PBGs for the interface states that are generated by the two PCs. The region, which is surrounded by red dashed lines, represents the frequency range of the fifth PBG of $PC_1$, $Im(Z_1) > 0$. The region, which is surrounded by blue solid lines, represents the frequency ranges of the fourth PBG of $PC_2$ with changing $a_{y2}$, $Im(Z_2) < 0$. Furthermore, the blue dashed lines indicate $a_{y2}/a_x = 0.7$. The region of black vertical lines represents the region for $Im(Z_1) > 0$ and $Im(Z_2) < 0$, when $a_{y1}$ is fixed, and $a_{y2}$ is changed. For these parameters, a certain frequency, which exists in any vertical line, can satisfy Eq. (1). We can obtain the frequency for the interface state in the same way as we obtained it from the blue dotted line in Fig. 2(a); (g) when $a_{y2} = 0.7a$ and $k_y = 3\pi/(5.6a)$, the projected band structures intercepted from Fig. 2(d) along the $k_x$ direction. The ordinal numbers of the PBGs are labeled with yellow.

According to Eq. (2), the surface impedance correlates with the Zak phase, and interface states can be confirmed according to Zak phases. As shown in Fig. 3(a)–3(d), when $k_y = 3\pi/(5.6a)$ and $a_y \in [0.45a, 0.705a)$, the Zak phases of the second and the third bands are $\pi$. Furthermore, when $a_y \in (0.705a, 1.4a]$, the Zak phases of the second and the third bands are 0, i.e., the second and the third bands are inverted when $a_y = 0.705a$. However, the Zak phase of the first (fourth) band is always $\pi$ (0). According to Eq. (2), although the Zak phases of the second and the third bands change as $a_y$ changes (see Fig. 3(b) and 3(c)), Im(Z) of the fourth PBG is still negative, and Im(Z) of the fifth PBG is still positive. In other words, whether the Zak phases of the second and the third bands of $PC_2$ change has no effect on the generation of interface states. By changing $a_{y1}$ and $a_{y2}$, the positions of PBGs of the PCs on both sides can be adjusted. Hence, the frequency range of the fifth PBG of $PC_1$ and the frequency range of the fourth PBG of $PC_2$ can overlap, and then Eq. (1) can be satisfied and the interface states can be generated.

According to the above analysis, by adjusting $a_{y2} \in [0.45a, 1.15a]$ with $a_{y1} = 1.4a$ fixed, the topological properties of the interface states in the PCs and the retainability of the interface states of the changeable rectangular lattices are studied—see Fig. 3(f). The region, which is surrounded by red dashed lines, represents the frequency range for Im($Z_1$) > 0 of the fifth PBG of $PC_1$, while the bands of $PC_1$ and Im($Z_1$) are shown in Fig. 3(e). The region, which is surrounded by blue solid lines, represents the frequency ranges for Im($Z_2$) < 0 of the fourth PBG of $PC_2$ with a changing $a_{y2}$. When $a_{y2} = 0.7a$, marked as the ends of the blue dashed lines, the bands of $PC_2$ and Im($Z_2$) are shown in Fig. 3(g). In the region of vertical lines in Fig. 3(f), when $a_{y1}$ and $a_{y2}$ are both fixed, there is a frequency for the corresponding vertical line (where the signs of Im(Z) of the fifth PBG of $PC_1$ and the fourth PBG of $PC_2$ are different) that can satisfy Eq. (1). As a result, an interface state can exist.

The situation that interface states can still exist, when the length-width ratio of the rectangular lattice of PC changes greatly, is called the retainability of the interface states in rectangular lattice PCs (referred to as retainability from here on). Although interface states can still exist, the frequencies of the interface states may change. Because of the retainability, when the frequency of incident light changes, the utilization of waveguide can still be realized with an interface state generated by adjusting $a_{y2}$. According to Fig. 3(f), at $k_y = 3\pi/(5.6a)$, $a_{y2} \in (0.460a, 1.136a)$, when interface states can be generated.

*3.2. Generation and description of the interface states with the same PBGs of the two PCs*

The interface states above are generated by utilizing the retainability of the interface states with the different PBGs of the two PCs and changing the rectangular lattice PCs. According to the above analysis of both Zak phases and surface impedances, by adjusting $a_y$ to change the corresponding Zak phase for each band, the third PBGs of $PC_1$ and $PC_2$ may be made to have different signs for Im($Z$). If their frequency ranges overlap, interface states can be generated within this overlapping frequency-range. The case that $a_{y1} = 0.9a$ ($\eta_1 = 0.9$), $a_{y2} = 0.45a$ ($\eta_2 = 0.45$) is studied to explore the feasibility to generate interface states with the same PBGs (the third PBGs of $PC_1$ and $PC_2$). The results are shown in Fig. 4.

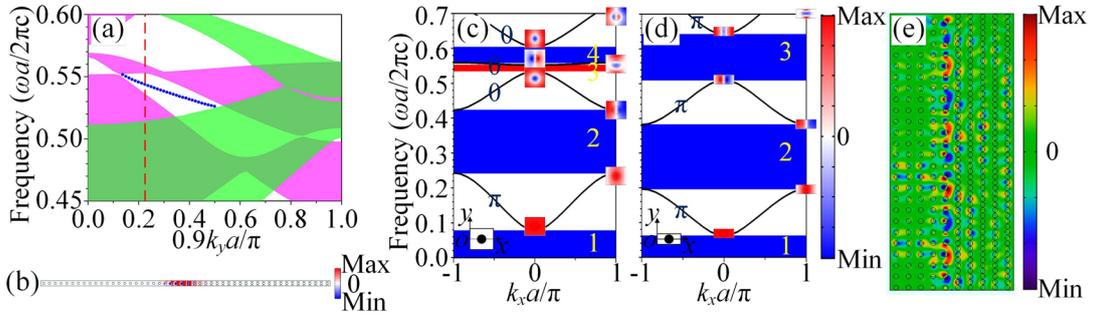

Fig. 4. (a) For $a_{y1} = 0.9a$ and $a_{y2} = 0.45a$, the projected band structures of the two PCs (the red dashed line indicates $k_y = \pi/(4a)$). The pink regions represent the passing bands of $PC_1$, and the light-green regions represent the passing bands of $PC_2$. The dark-green regions represent the common bands of the two PCs. The band marked with a blue dotted line represents the band of the interface states; (b) the eigenfield distribution of the interface state at $f = 0.5400c/a$ with $k_y = \pi/(4a)$ in the projected band structures (the numbers of dielectric rods in the $x$ direction for the regions of both $PC_1$ and $PC_2$ are 25); the projected band structures in the $k_x$ direction, eigenfield distributions, and

Zak phases at $k_y = \pi/(4a)$ of the two PCs: (c) PC$_1$; (d) PC$_2$. The colors of the PBGs represent the signs of Im(Z), with red indicating positive and blue indicating negative. The coordinates for calculating surface impedances and Zak phases are shown in the lower left corner of the diagrams, and the origins are in the center of the left boundaries of the cells. The ordinal numbers of the PBGs are labeled with yellow; (e) the field distribution of the generated interface state at $f = 0.5400c/a$ (the numbers of dielectric rods in the $x$ direction for the regions of both PC$_1$ and PC$_2$ are 5, and the numbers of dielectric rods in the $y$ direction for the regions of PC$_1$ and PC$_2$ are 25 and 50, respectively).

As shown by the blue dotted line in Fig. 4(a), there are indeed interface states in the common PBG. Fig. 4(b) shows the eigenfield distribution at $f = 0.5400c/a$ with $k_y = \pi/(4a)$ for the projected band structures. The eigenfield distribution is localized at the interface of the two PCs. According to Fig. 4(c) and 4(d), at $k_y = \pi/(4a)$, Im(Z) for the third PBG of PC$_1$ (PC$_2$) is positive (negative), and the frequency ranges of the third PBGs of the two PCs overlap. As a result, an interface state can be generated in this overlapped frequency range, which is consistent with the frequency (the same as that in Fig. 4(b)) satisfying Eq. (1) in Fig. 4(a). According to Fig. 4(e), an interface state exists at the interface of the two PCs at $f = 0.5400c/a$ (the same as that in Fig. 4(b)).

Similar to Fig. 3, to better realize the retainability of the interface states, both Zak phases and surface impedances are used to study the retainability of the interface states at $k_y = \pi/(4a)$ by changing the length-width ratio ($\eta_2$) of the rectangular lattice below. The results are shown in Fig. 5.

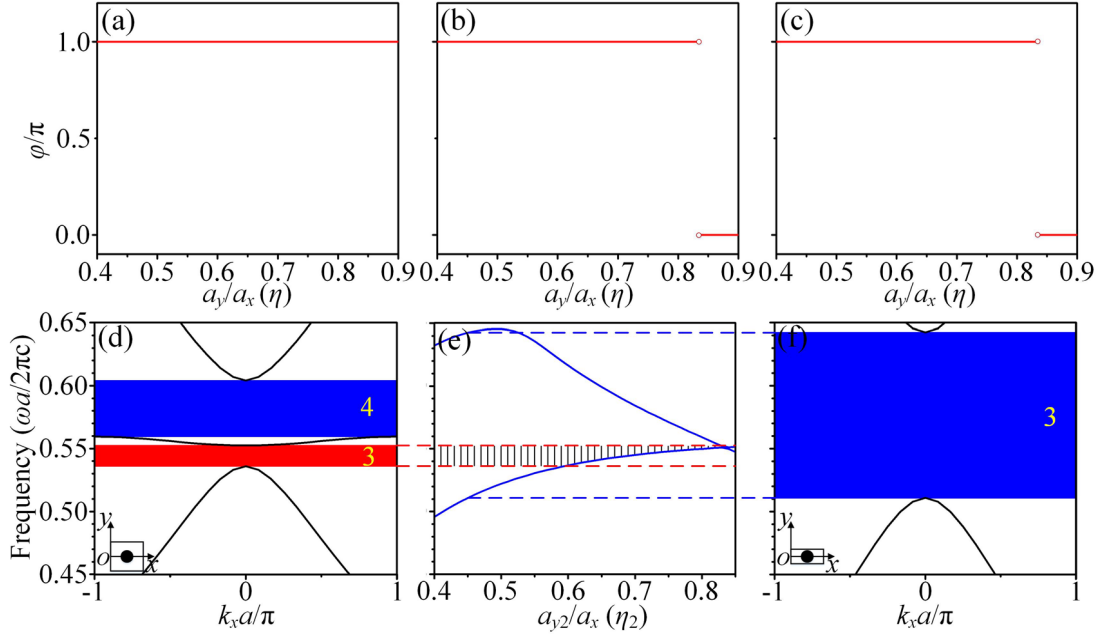

Fig. 5. (a)–(c) For $k_y = \pi/(4a)$ and $a_y \in [0.4a, 0.9a]$, the Zak phase is shown for: (a) the first band, (b) the second band, (c) the third band; (d) when $a_{y1} = 0.9a$ and $k_y = \pi/(4a)$, the projected band structures intercepted from Fig. 4(c) along the $k_x$ direction. The ordinal numbers of the PBGs are labeled with yellow; (e) when $a_{y1} = 0.9a$, $a_{y2} \in [0.4a, 0.85a]$ and $k_y = \pi/(4a)$, the schematic diagram of the frequency ranges of the corresponding PBGs for the interface states that are generated by the two PCs. The region, which is surrounded by red dashed lines, represents the frequency range of the third PBG of PC$_1$, Im($Z_1$) > 0. The region, which is surrounded by blue solid lines, represents the frequency ranges of the third PBG of PC$_2$ with changing $a_{y2}$, Im($Z_2$) < 0. Furthermore, the blue dashed lines indicate $a_{y2}/a_x = 0.45$. The region of black vertical lines represents the region for Im($Z_1$) > 0 and Im($Z_2$) < 0, when $a_{y1}$ is fixed and $a_{y2}$ is changed. For these parameters, a certain frequency, which exists in any vertical line, can satisfy Eq. (1). We can obtain the frequency for the interface state in the same way as we obtained it from the blue dotted line in Fig. 2(a); (f) when $a_{y2} = 0.45a$ and $k_y = \pi/(4a)$, the projected band structures intercepted from Fig. 4(d) along the $k_x$ direction. The ordinal numbers of the PBGs are labeled with yellow.

As shown in Fig. 5(a)–5(c), when $k_y = \pi/(4a)$ and $a_y \in [0.4a, 0.835a)$, the Zak phases of the second and the third bands are π. Furthermore, when $a_y \in (0.835a, 0.9a]$, the Zak phases of the second and the third bands are 0 (the second and the third bands are inverted when $a_y = 0.835a$), and the Zak phase of the first band is always π. According to the above analysis, by adjusting $a_{y2} \in [0.4a, 0.85a]$ with $a_{y1} = 0.9a$ fixed, the topological properties of the interface states of the PCs and the retainability of the interface states of the changing rectangular lattice are studied—see Fig. 5(e). The region, which is surrounded by red dashed lines, represents the frequency range for $Im(Z_1) > 0$ of the third PBG of $PC_1$, and the bands of $PC_1$ and $Im(Z_1)$ are shown in Fig. 5(d). The region, which is surrounded by blue solid lines, represents the frequency ranges for $Im(Z_2) < 0$ of the third PBG of $PC_2$ with a changing $a_{y2}$. When $a_{y2} = 0.45a$, marked as the ends of the blue dashed lines, the bands of $PC_2$ and $Im(Z_2)$ are shown in Fig. 5(f). In the region of vertical lines in Fig. 5(e), when $a_{y1}$ and $a_{y2}$ are both fixed, there is a frequency for the corresponding vertical line (where the signs of $Im(Z)$ of the third PBGs of the two PCs are different) that can satisfy Eq. (1). This means that an interface state can exist.

*3.3. The comparison between the interface states in 3.1 and 3.2*

No matter whether the second and the third bands of $PC_2$ are inverted (i.e. the Zak phases are changed) with the different PBGs, the interface states can be generated (see Fig. 3), as long as the frequency ranges of the fifth PBG of $PC_1$ and the fourth PBG of $PC_2$ overlap. This means that the retainability is related to the positions of the PBGs. On the other hand, the interface states can only be generated when the second and the third bands of $PC_2$ are not inverted ($a_{y2} < 0.835$, $Im(Z_1) > 0$, $Im(Z_2) < 0$) with the same PBGs (see Fig. 5). Interface states cannot be generated, however, once the two bands are inverted ($a_{y2} > 0.835$, $Im(Z_1) > 0$, $Im(Z_2) > 0$). As a result, the retainability is related to the Zak phases of the bands.

It should be emphasized the fact that $a_{y2} \in [0.4a, 0.835a)$ (at $k_y = π/(4a)$), when interface states exist with the same PBGs, is related to the characteristic of $PC_2$ (the Zak phases of the bands of $PC_2$), where the second and the third bands are inverted when $a_y = 0.835a$. However, it can be found the fact that $a_{y2} \in (0.460a, 1.136a)$

(at $k_y = 3\pi/(5.6a)$), when interface states exist with the different PBGs, is related to the characteristic of PC$_1$ and PC$_2$ (the positions of the PBGs of the two PCs). Hence, the range of $a_{y2}$ can be expanded, when interface states exist with the different PBGs, by changing the structure of PC$_1$ (the retainability is adjustable). The retainability with the same PBGs is not adjustable because of band inversion, which leads to the change of the topological properties of the bands (the change happens when $a_{y2}/a_x = 0.835$ in Fig. 5(e)).

The above conclusions are related to the relationship between the impedances, the widths, the positions of PBGs, and the lattice constants. In brief, the retainability of the interface states depends on the changing rule of the Zak phases of the bands and the PBGs with the lattice constant in rectangular lattice PCs. A more detailed analysis of the characteristics of the interface states in the rectangular lattice PCs, with different materials, shapes, and sizes of dielectric rods, shows that these conclusions are also applicable (see the **Supplementary Material**).

As shown above, if we want to generate interface states, we can change not only the materials, the shapes and the sizes of the dielectric rods, but also the length-width ratio of the lattice. Different from the square lattice, the honeycomb lattice or the triangular lattice, the variability of the rectangular lattice helps us to utilize the interface states more conveniently because the frequency range of the interface states can be easily changed by adjusting the length-width ratio of the lattice.

## 4. Conclusion

Two PCs, with different rectangular lattices but identical materials, shapes and sizes of dielectric rods on both sides of the interface, are used to construct interface. This enables the easy generation of interface states. Then, the mechanism of generating interface states related to Zak phases and surface impedances is investigated. When interface states are generated in rectangular lattice PCs, the retainability of the interface states is investigated for the topological properties of the Zak phase at the selected $k_y$, under the circumstance that the length-width ratio of the rectangular lattice changes (fixing the length-width ratio ($\eta_1$) of a rectangular lattice of $PC_1$ and changing the length-width ratio ($\eta_2$) of the other rectangular lattice of $PC_2$). It is found that the retainability is related to the positions of the PBGs or the Zak phases of the bands. These observations verify the feasibility of generating interface states by only changing the length-width ratio of the rectangular lattices and help improve the understanding of the interface states in the rectangular lattice PCs made of the same materials, shapes and sizes of dielectric rods. These researches may do help to the understanding of the relationship between geometry and interface states, and enable us to adjust the lattice size to realize optimized transmission in waveguides based on the generation of interface states.

**Acknowledgements**

This work was supported by the National Natural Science Foundation of China (Grant Nos. 61405058 and 62075059), the Natural Science Foundation of Hunan Province (Grant Nos. 2017JJ2048 and 2020JJ4161), and the Fundamental Research Funds for the Central Universities (Grant No. 531118040112). The authors acknowledge Professor J. Q. Liu for software sponsorship.

**Supplementary Material:**

**The characteristics of the interface states in the rectangular lattice PCs with different materials, shapes, and sizes of dielectric rods**

It is assumed that the materials of the square dielectric rods of the two PCs are germanium ($\varepsilon_r = 16$), with a side length $d = 0.32\,\mu m$. The structures of the combined PCs are: Structure 1: $a_{y1} = 1.4a$ ($\eta_1 = 1.4$), $a_{y2} = 0.7a$ ($\eta_2 = 0.7$); Structure 2: $a_{y1} = 0.9a$ ($\eta_1 = 0.9$), $a_{y2} = 0.45a$ ($\eta_2 = 0.45$). We investigate whether the two structures can, respectively, generate interface states with the different or the same PBGs of the two PCs—see Fig. S1. The below results are all for TM polarization.

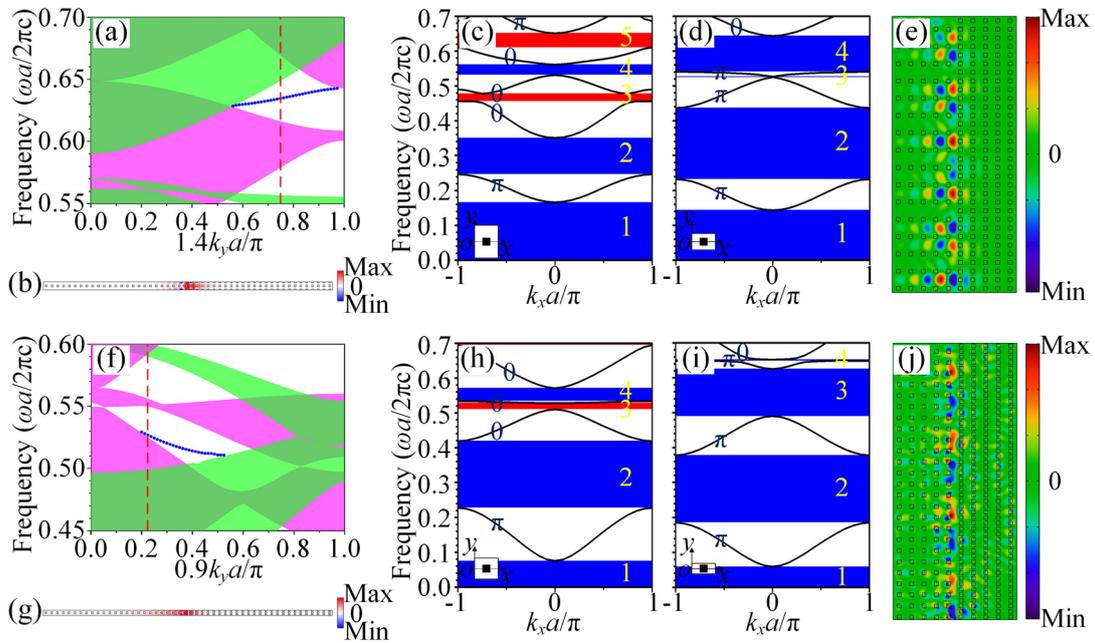

Fig. S1. (a), (f) The projected band structures of the PCs for Structures 1 and 2. The red dashed lines indicate $k_y = 3\pi/(5.6a)$ and $\pi/(4a)$ respectively, and the corresponding frequencies of the interface states are $f = 0.6343c/a$ and $0.5266c/a$ respectively. The pink regions represent the passing bands of $PC_1$, and the light-green regions represent the passing bands of $PC_2$. The dark-green regions represent the common bands of the two PCs. The bands marked with blue dotted lines represent the bands of the interface states; (b), (g) the eigenfield distributions of the interface states at $f = 0.6343c/a$ ($k_y = 3\pi/(5.6a)$) and $f = 0.5266c/a$ ($k_y = \pi/(4a)$) in the projected band structures, respectively, of Structures 1 and 2; the projected band structures along the $k_x$ direction and the Zak phases at $k_y = 3\pi/(5.6a)$ of the two PCs in Structure 1: (c) PC$_1$; (d) PC$_2$. The colors of the PBGs represent the signs of Im(Z), with red indicating positive and blue indicating negative. The ordinal numbers of the PBGs are labeled with yellow; (e) the field distribution of the generated interface state, at $f = 0.6343c/a$ in Structure 1; the projected band structures along the $k_x$ direction and the Zak phases at $k_y = \pi/(4a)$ of the two PCs in Structure 2: (h) PC$_1$, (i) PC$_2$. The colors of the PBGs represent the signs of Im(Z), with red indicating positive and blue indicating negative. The ordinal numbers of the PBGs are labeled with yellow; (j) the field distribution of the generated interface state, at $f = 0.5266c/a$ in Structure 2.

According to Fig. S1(a) and S1(f), both structures can generate interface states. According to Fig. S1(b) and S1(g), the eigenfield distributions are localized at the interfaces of the two PCs. Similar to the analysis of surface impedances and Zak phases in Fig. 2(c) and 2(d), in Fig. S1(c) and S1(d), there is a frequency that satisfies Eq. (1) in the fifth PBG of PC$_1$ and the fourth PBG of PC$_2$ in Structure 1. Hence, an interface state exists, with the different PBGs of the two PCs [SF1–SF5]. In Fig. S1(e), an interface state exists at the interface between the two PCs of Structure 1, at $f = 0.6343c/a$. Similar to the analysis of surface impedances and Zak phases in Fig. 4(c) and 4(d), in Fig. S1(h) and S1(i), there is a frequency that satisfies Eq. (1) in the third PBGs of PC$_1$ and PC$_2$ in Structure 2. Hence, an interface state exists, with the same PBGs of the two PCs [SF1–SF5]. In Fig. S1(j), an interface state exists at the interface between the two PCs of Structure 2, at $f = 0.5266c/a$.

Similar to Fig. 3 and 5, by changing the length/width ratio of the rectangular lattice ($\eta_2$) of PC$_2$, the retainability of the two kinds of interface states are studied according to Zak phases and surface impedances with the different PBGs (see Fig. S2, $k_y = 3\pi/(5.6a)$) and the same PBGs (see Fig. S3, $k_y = \pi/(4a)$) of the two PCs. This is done with the aim to research the factors of the generation of interface states.

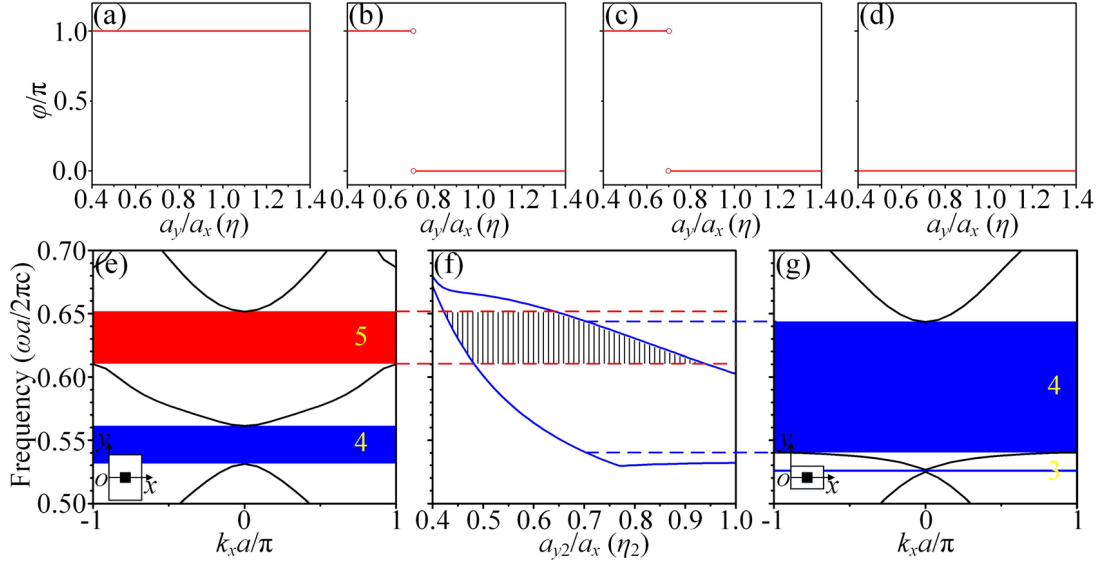

Fig. S2. (a)–(d) For $k_y = 3\pi/(5.6a)$ and $a_y \in [0.4a, 1.4a]$, the Zak phase is shown for: (a) the first band, (b) the second band, (c) the third band, (d) the fourth band; (e) when $a_{y1} = 1.4a$ and $k_y = 3\pi/(5.6a)$, the projected band structures intercepted from Fig. S1(c) along the $k_x$ direction. The ordinal numbers of the PBGs are labeled with yellow; (f) when $a_{y1} = 1.4a$, $a_{y2} \in [0.4a, a]$ and $k_y = 3\pi/(5.6a)$, the schematic diagram of the frequency ranges of the corresponding PBGs for the interface states that are generated by the two PCs. The region, which is surrounded by red dashed lines, represents the frequency range of the fifth PBG of $PC_1$, $Im(Z_1) > 0$. The region, which is surrounded by blue solid lines, represents the frequency ranges of the fourth PBG of $PC_2$ with changing $a_{y2}$, $Im(Z_2) < 0$. Furthermore, the blue dashed lines indicate $a_{y2}/a_x = 0.7$. The region of black vertical lines represents the region for $Im(Z_1) > 0$ and $Im(Z_2) < 0$, when $a_{y1}$ is fixed and $a_{y2}$ is changed. For these parameters, a certain frequency, which exists in any vertical line, can satisfy Eq. (1). We can obtain the frequency for the interface state in the same way as we obtained it from the blue dotted line in Fig. 2(a); (g) when $a_{y2} = 0.7a$ and $k_y = 3\pi/(5.6a)$, the projected band structures intercepted from Fig. S1(d) along the $k_x$ direction. The ordinal numbers of the PBGs are labeled with yellow.

As shown in Fig. S2(a)–S2(d), when $k_y = 3\pi/(5.6a)$ and $a_y \in [0.4a, 0.701a)$, the Zak phases of the second and the third bands are π, and when $a_y \in (0.701a, 1.4a]$, the Zak phases of the second and the third bands are 0. This means, the second and the third bands are inverted when $a_y = 0.701a$. The Zak phase of the first (fourth) band is always π (0), which is similar to the change rule in Fig. 3(a)–3(d). Hence, we can analyze the retainability of interface states in a similar way—see Fig. S2(f). The region, which is surrounded by red dashed lines, represents the frequency range for $Im(Z_1) > 0$ of the fifth PBG of $PC_1$, while the bands of $PC_1$ and $Im(Z_1)$ are shown in Fig. S2(e). The region, which is surrounded by blue solid lines, represents the frequency ranges for $Im(Z_2) < 0$ of the fourth PBG of $PC_2$ with a changing $a_{y2}$. For $a_{y2} = 0.7a$, i.e., the ends of the blue dashed lines, the bands of $PC_2$ and $Im(Z_2)$ are shown in Fig. S2(g). In the region with vertical lines in Fig. S2(f), when $a_{y1}$ and $a_{y2}$ are fixed, there is a frequency for the corresponding vertical line that can satisfy Eq. (1). This means that interface states can exist. According to Fig. S2(f), at $k_y = 3\pi/(5.6a)$, $a_{y2} \in (0.421a, 0.936a)$ when interface states are generated with the different PBGs of the two PCs, and the retainability is related to the positions of the PBGs.

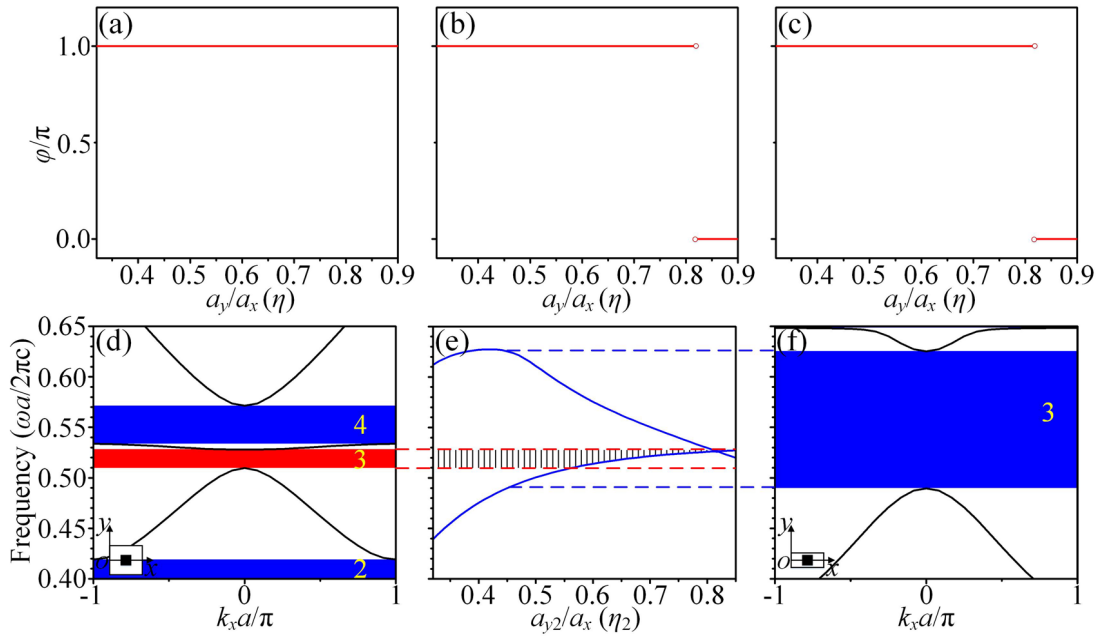

Fig. S3. (a)–(c) For $k_y = \pi/(4a)$ and $a_y \in [0.32a, 0.9a]$, the Zak phase is shown for: (a) the first band, (b) the second band, (c) the third band; (d) when $a_{y1} = 0.9a$ and $k_y = \pi/(4a)$, the projected band structures intercepted from Fig.

S1(h) along the $k_x$ direction. The ordinal numbers of the PBGs are labeled with yellow; (e) when $a_{y1} = 0.9a$, $a_{y2} \in$ [0.32$a$, 0.85$a$] and $k_y = \pi/(4a)$, the schematic diagram of the frequency ranges of the corresponding PBGs for the interface states that are generated by the two PCs. The region, which is surrounded by red dashed lines, represents the frequency range of the third PBG of PC$_1$, Im($Z_1$) > 0. The region, which is surrounded by blue solid lines, represents the frequency ranges of the third PBG of PC$_2$ with changing $a_{y2}$, Im($Z_2$) < 0. Furthermore, the blue dashed lines indicate $a_{y2}/a_x = 0.45$. The region of black vertical lines represents the region for Im($Z_1$) > 0 and Im($Z_2$) < 0, when $a_{y1}$ is fixed and $a_{y2}$ is changed. For these parameters, a certain frequency, which exists in any vertical line, can satisfy Eq. (1). We can obtain the frequency for the interface state in the same way as we obtained it from the blue dotted line in Fig. 2(a); (f) when $a_{y2} = 0.45a$ and $k_y = \pi/(4a)$, the projected band structures intercepted from Fig. S1(i) along the $k_x$ direction. The ordinal numbers of the PBGs are labeled with yellow.

As shown in Fig. S3(a)–S3(c), when $k_y = \pi/(4a)$ and $a_y \in [0.32a, 0.818a)$, the Zak phases of the second and the third bands are $\pi$, and when $a_y \in (0.818a, 0.9a]$, the Zak phases of the second and the third bands are 0. This means, the second and the third bands are inverted when $a_y = 0.818a$. The Zak phase of the first band is always $\pi$, which is similar to the change rule in Fig. 5(a)–5(c). Hence, we can analyze the retainability of interface states in a similar way—see Fig. S3(e). The region, which is surrounded by red dashed lines, represents the frequency range for Im($Z_1$) > 0 of the third PBG of PC$_1$, while the bands of PC$_1$ and Im($Z_1$) are shown in Fig. S3(d). The region, which is surrounded by blue solid lines, represents the frequency ranges for Im($Z_2$) < 0 of the third PBG of PC$_2$ with a changing $a_{y2}$. For $a_{y2} = 0.45a$, i.e., the ends of the blue dashed lines, the bands of PC$_2$ and Im($Z_2$) are shown in Fig. S3(f). In the region with vertical lines in Fig. S3(e), when $a_{y1}$ and $a_{y2}$ are fixed, there is a frequency for the corresponding vertical line that can satisfy Eq. (1). This means that interface states can exist. According to Fig. S3(e), at $k_y = \pi/(4a)$, $a_{y2} \in [0.32a, 0.818a)$, when interface states are generated with the same PBGs of the two PCs, and the retainability is related to the Zak phases of the bands.

According to the above analysis for $\varepsilon_r$ = 16 and $d$ = 0.32μm, and after comparing Fig. S2(f) with Fig. S3(e), it can be found that $a_{y2} \in (0.421a, 0.936a)$ at $k_y = 3\pi/(5.6a)$, when interface states exist with the different PBGs of the two PCs, and the retainability is related to the positions of PBGs. On the other hand, $a_{y2} \in [0.32a, 0.818a)$ at $k_y = \pi/(4a)$, when interface states exist with the same PBGs of the two PCs, and the retainability is related to the Zak phases of the bands. Similar to the Text, the retainability of interface states with the different PBGs can be adjusted by changing $PC_1$. However, the retainability of interface states with the same PBGs cannot be adjusted. The above conclusions about the characteristics of the interface states in the rectangular lattice PCs are similar to those in the Text.